\documentclass[aip,
rsi,%
 amsmath,amssymb,
 reprint,%
]{revtex4-1} 

\usepackage{amsmath}  
\usepackage{amsfonts} 
\usepackage{siunitx} 
\usepackage{verbatim}
\usepackage{color}
\usepackage[pdftex]{graphicx}
\usepackage{epstopdf} 
\usepackage{hyperref}
\usepackage[english]{babel}
\usepackage{blindtext}

\begin{document}
\preprint{AIP/123-QED}

\title{Quantitative Stirling Cycle Measurements: P-V Diagram and Refrigeration}

\author{Y. J. Lu}
\author{Hiroko Nakahara}

\author{J. S. Bobowski}
\email{jake.bobowski@ubc.ca} 
\affiliation{Department of Physics, University of British Columbia, Kelowna, British Columbia, Canada V1V 1V7}


\date{\today}

\maketitle 

\section{Introduction}\label{sec:intro}

This paper describes simple modifications to a demonstration Stirling engine that allows one to make quantitative measurements of the Stirling cycle.  First, we describe measurements of the $P$-$V$ diagram using an inexpensive pressure sensor and a common photogate.  Then, as a supplement, the engine was run as a refrigerator by using a rotary tool to turn the flywheel of the engine. A small, but measurable, temperature difference developed across the body of the engine.

Stirling engines have long been used as in-class demonstrations and, \cite{TPT:1982, TPT:1990} in some cases, $P$-$V$ diagrams have be constructed using a commercial apparatus.\cite{PhysEduc:1994} It is also worth noting that, despite the fact that Stirling's original patent was filed over 200 year ago, the Stirling cycle is still being actively studied.\cite{AJP:1984, AJP:2017}  The main objective of our work is to make an accurate measurement of the $P$-$V$ diagram of a high-quality Stirling engine using inexpensive materials and a data acquisition system that is commonly found in undergraduate laboratories. 

This work was inspired by an earlier measurement of a Stirling engine $P$-$V$ diagram by Nakahara.\cite{Nakahara:2009}  In that project, the $P$-$V$ diagram was measured using Vernier's Logger Pro data acquisition system.  Pressure data were acquired at a rate of 10 samples per second using Vernier's pressure sensor.  The gas volume was measured by mounting a reflector onto the piston of the engine and using an infrared (IR) transmitter/receiver to track the piston's position.  During operation, the calibrated output voltage of the IR sensor was monitored using Vernier's differential voltage probe.  

The advantage of Nakahara's scheme is that it produced a real-time measurement of the $P$-$V$ diagram.  There are, however, a number of limitations and disadvantageous to this scheme.  One is that the pressure sensor does not have the sensitivity to to precisely measure the small pressure changes associated with the demonstration Stirling engine.  Second, the time between pressure measurements is long compared to the period of the Stirling cycle.  Third, the calibration of the IR sensor is nonlinear such that the measurement sensitivity of the piston position is low when the reflector is furthest from the IR transmitter/detector.  Finally, to mount the IR sensor, a new top plate for the Stirling engine had to be machined.  In this project, we developed a measurement scheme to overcome these limitations which allowed us to produce high-quality pressure and volume measurements when using the same Striling engine as Nakahara.

When covering the topic of heat engines in an introductory thermodynamics course, one typically first describes the Carnot cycle.\cite{Schroeder:1999}  In the Carnot cycle, the isothermal compression of an ideal gas results in heat $Q_1$ being released to a cold thermal reservoir at temperature $T_1$ whereas an isothermal expansion causes the gas to absorb heat $Q_2$ from a hot thermal reservoir at temperature $T_2>T_1$.  The efficiency $\varepsilon$ of the cycle is determined by the fraction of the heat removed from the hot reservoir that is converted into useful work $W$:
\begin{equation}
\varepsilon=\frac{W}{Q_2}=1-\frac{T_1}{T_2}\label{eq:eff}
\end{equation}   
In addition to the isothermal processes, the Carnot cycle requires an adiabatic compression and expansion to take the gas from $T_1$ to $T_2$ and $T_2$ to $T_1$, respectively.  The Carnot cycle is conceptually simple because there is no heat exchanged between the ideal gas and the thermal reservoirs during the adiabatic processes.  The cycle is, however, difficult to implement in practice precisely because one cannot easily isolate the gas from the thermal reservoirs.

\begin{figure}[t]
\centering{
\includegraphics[keepaspectratio, width=.9\columnwidth]{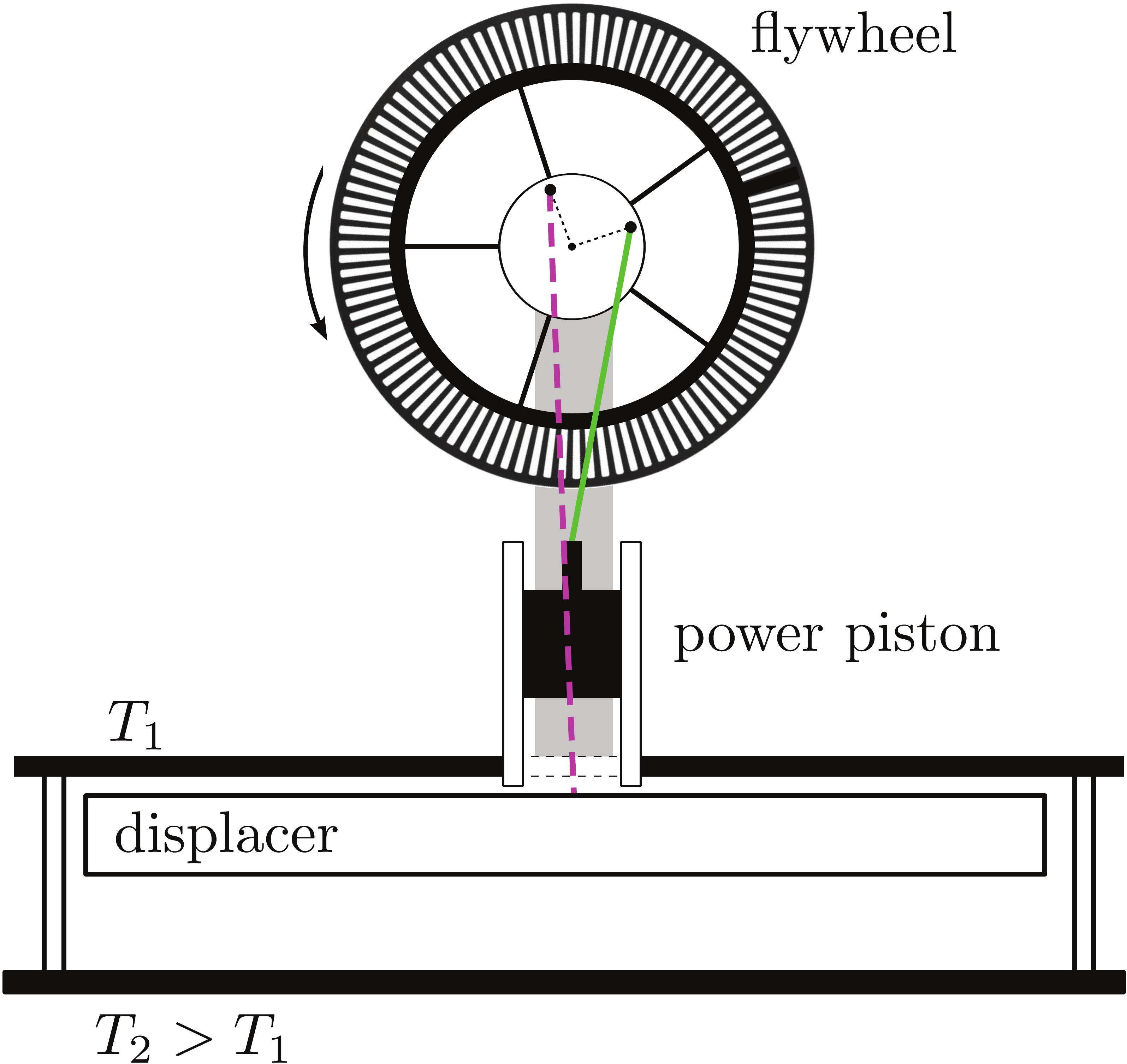}}
\caption{\label{fig:schematic}Schematic diagram of the displacer-style Stirling engine.  The power piston and displacer move \SI{90}{\degree} out of phase with one another.  A temperature difference with $T_2>T_1$ will result in a counterclockwise rotation of the flywheel.}
\end{figure}
The Stirling cycle, on the other hand, is easier to implement but more challenging conceptually since it requires the idea of a regenerator.  In the Stirling cycle, the adiabatic processes of the Carnot cycle are replaced with isochoric (constant volume) processes.  The isochoric parts of the cycle do no work, however, the gas must absorb (release) heat as its pressure and temperature increases (decreases) at constant volume.  This heat exchange occurs via a regenerator through which the gas is forced to pass.  The regerenator is typically a porous mesh with a large heat capacity.  It can absorb or release a small amount of heat without a substantial change to its temperature.  An analysis of the ideal Stirling cycle shows that it has the same efficiency, given by Eq.~\ref{eq:eff}, as the Carnot cycle.  

The Stirling cycle is more practical because the constant volume processes can be implemented in a number of different ways. The Stirling engine used in our project is made by Kontax Stirling Engines.\cite{Kontax:2018}  The main elements of the engine, shown schematically in Fig.\ref{fig:schematic}, are a power piston that sets the volume of the working gas, a displacer that is used to move the gas between the hot and cold regions of the engine, and a flywheel.  The positions of the piston and displacer are set by connecting rods that are attached to a disk centered on the flywheel.  As shown in the figure, the connecting rods are attached such that the piston and displacer positions are \SI{90}{\degree} out of phase with one another.  In Fig.~\ref{fig:schematic}, a small counterclockwise rotation of the flywheel would result in an (approximately) isothermal expansion of the gas while it is in contact with the hot reservoir.  This would be followed by an (approximately) isochoric process that would decrease the gas temperature from $T_2$ to $T_1$.  

\begin{figure}[t]
\centering{
\includegraphics[keepaspectratio, width=\columnwidth]{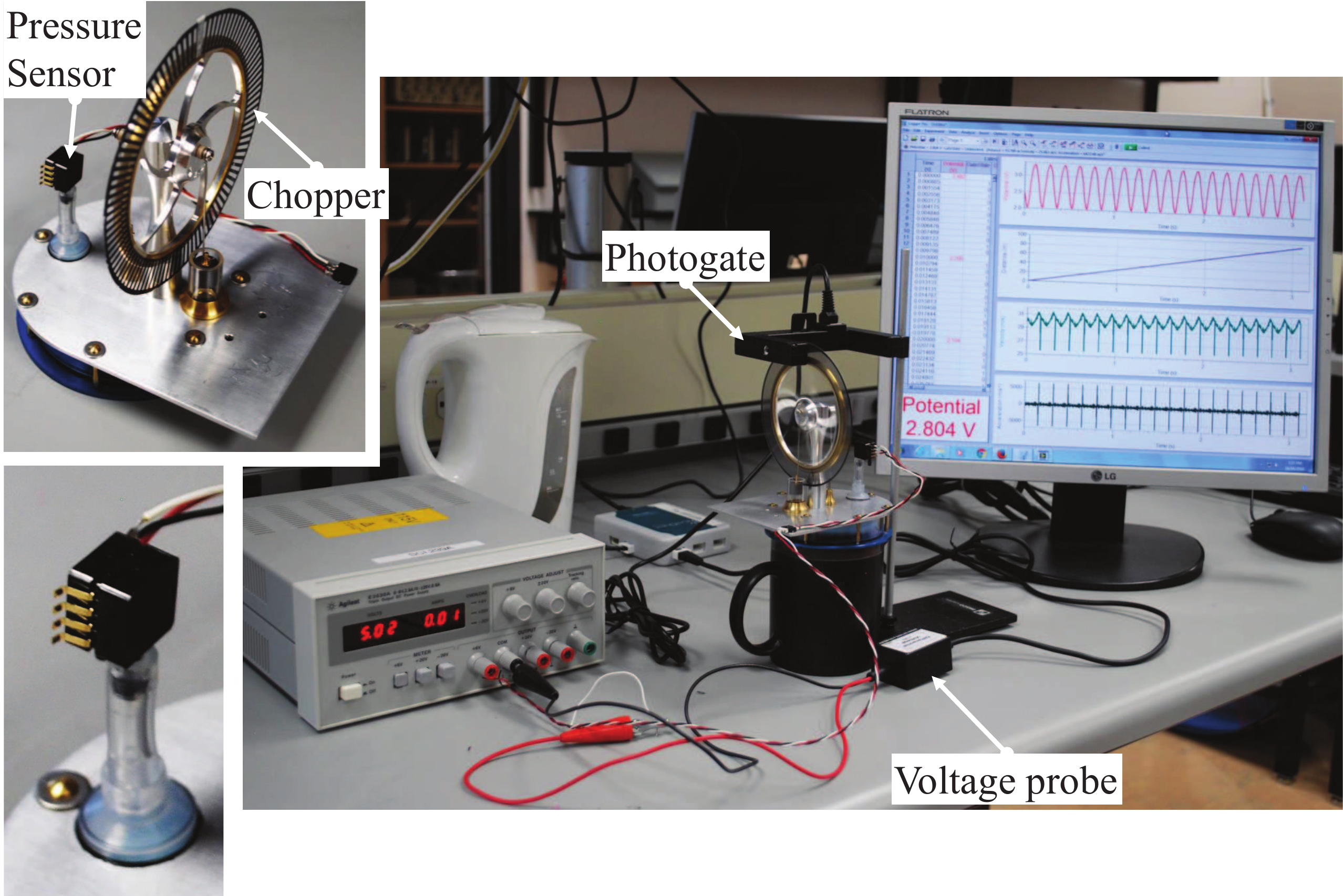}}
\caption{\label{fig:PVapparatus}Photograph of the setup to measure the Striling cycle $P$-$V$ diagram.  A photogate and chopper are used to sense the position of the power piston while a pressure sensor and voltage probe monitor the gas pressure.}
\end{figure}
This displacer-style Stirling engine does not have a porous regenerator through which the gas passes during the constant volume processes.  However, the surface of the displacer and the cylindrical wall separating the hot and cold thermal reservoirs together act as a crude regenerator.  With such a poor regenerator, the work output per cycle will be low and the engine's efficiency is expected to be much less than that of the ideal Stirling cycle.

\section{P-V diagram}\label{sec:PV}

Figure~\ref{fig:PVapparatus} shows the experimental setup used to simultaneously measure the pressure and volume of the displacer-type Stirling engine.  A list of the all of the equipment and materials used in our experiments is given in Appendix~\ref{app:parts}. Where appropriate, possible vendors are suggested and estimates of costs are given.

\subsection{Pressure measurement}\label{sec:pressure}

\begin{figure}[t]
\centering{\includegraphics[width=0.94\columnwidth]{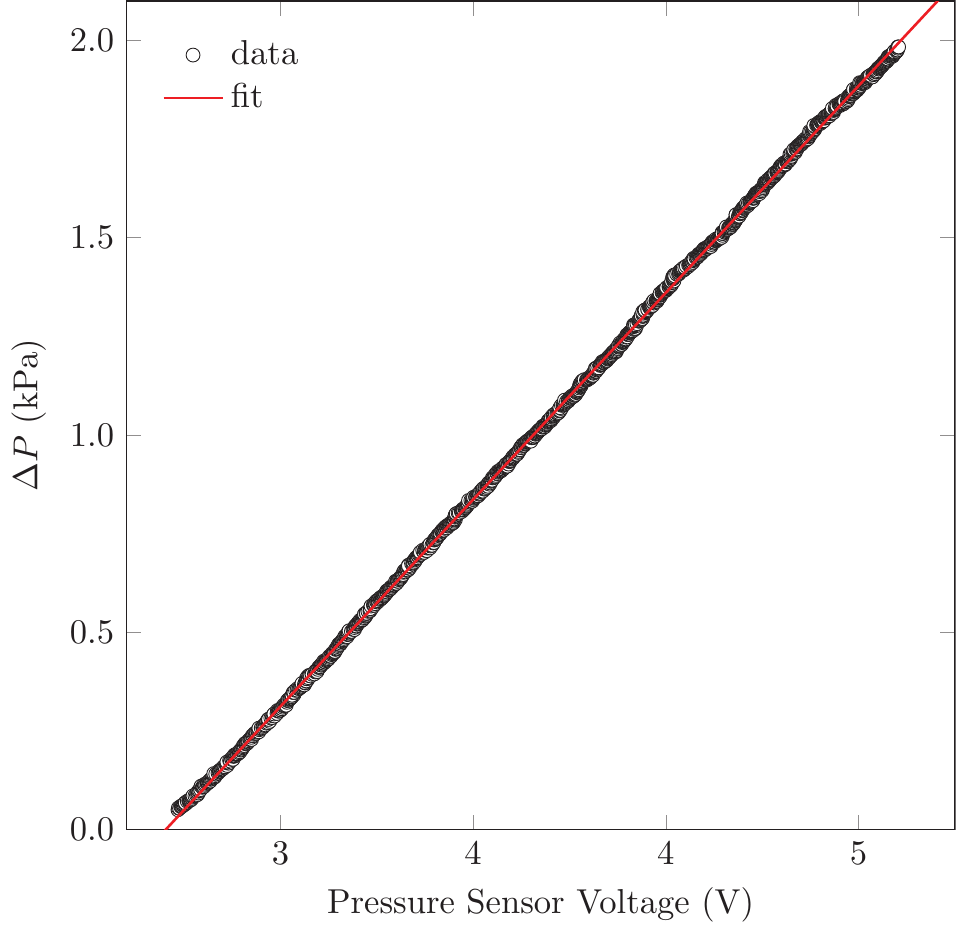}}
\caption{\label{fig:calibration}Pressure sensor calibration data for the MPXV7002 pressure sensor.  The measured temperature change of a fixed volume of air was converted to a pressure difference while simultaneously recording the voltage output of the pressure sensor.}
\end{figure}
To monitor the pressure change of the gas in the Stirling engine, an inexpensive differential pressure sensor, having the optimal sensitivity for our application, was purchased.  From Nakahara's previous work, the pressure change was expected to be just over \SI{1}{\kilo\pascal}.  Therefore, we purchase the MPXV7002 pressure sensor manufactured by NXP Semiconductors which has a voltage output that varies from \SI{0.5}{} to \SI{4.5}{\volt} for pressure changes from \SI{-2}{} to \SI[retain-explicit-plus]{+2}{\kilo\pascal}.  The datasheet supplied by NXP Semiconductors states that the typical calibration of the output is given by:
\begin{equation}
\Delta P=\left(V_\mathrm{out}-\SI{2.5}{\volt}\right)\times\SI{1}{\kilo\pascal/\volt}\label{eq:NPX}
\end{equation} 
where $\Delta P$ is the pressure difference across the sensor in kilopascals and $V_\mathrm{out}$ is the sensor output voltage in volts. (Equation~\ref{eq:NPX} assumes that the sensor is powered by \SI[retain-explicit-plus]{+5}{\volt}.)\cite{NXP:2017}

To precisely calibrate our specific pressure sensor, we developed a simple calibration apparatus.  Rubber tubing was used to connect the pressure sensor to one end of a length of standard 1/4-inch copper tubing. (The copper tubing was approximately \SI{1}{\meter} long, but the precise length is not critical and does not need to be known.)  A type-T thermocouple was then inserted into the opposite end until its tip was approximately in the center of the tube.  Finally, the open end of the copper tube (through which the thermocouple was fed) was sealed using J-B Weld epoxy.  The temperature of the copper tube was then very slowly raised above room temperature by about \SI{6}{\celsius} using a T-shirt heat press.  The heat press was plugged into a variac which was used to drop the supply voltage from \SI{110}{} to \SI{15}{Vrms} so as to limit the temperature increase.  The change in pressure of the air trapped in the copper tube is related to the temperature change via:
\begin{equation}
\Delta P=P\frac{\Delta T}{T}
\end{equation}
where $P$ and $T$ are the ambient pressure and temperature of the room, respectively.  The calibration data obtained are shown in Fig.~\ref{fig:calibration}.  A linear fit to the data yielded the following relationship between $\Delta P$ of the air and the pressure sensor output voltage:
\begin{equation}
\Delta P=\left(1.0485 V_\mathrm{out}-\SI{2.833}{\volt}\right)\times\SI{1}{\kilo\pascal/\volt}
\end{equation}
which is in reasonable agreement with manufacturer specifications.

To connect the pressure sensor to the Stirling engine, a hole was drilled through the engine's aluminum top plate to accommodate the tip of a Luer lock syringe.  The syringe was epoxied in place using J-B Weld.  As shown in Fig.~\ref{fig:PVapparatus}, the pressure sensor was then connected to the syringe tip using tight-fitting rubber tubing.  The figure shows a modified top plate, however, all that is needed is an appropriately-sized hole in the original plate.  We used a modified top plate only because it was left over from Nakahara's earlier project.  During operation, the pressure sensor's output voltage was recorded at a rate of \SI{200}{samples/\second} using Vernier's Logger Pro software and differential voltage probe.

\subsection{Volume measurement}\label{sec:volume}

To calculate the change in gas volume, the height of the Stirling engine's piston was measured.  A circular chopper, shown in Fig.~\ref{fig:chopper} of Appendix~\ref{app:chopper}, and a photogate were used to track the rotation of the engine's flywheel.  The chopper was designed to have 100 uniform dark and transparent intervals.  One of the transparent intervals was intentionally blacked out.  When the chopper was attached to the flywheel, as shown in Fig.~\ref{fig:PVapparatus}, the blackened segment was positioned such that it was at the top of the flywheel when the piston was at its maximum height.  This arrangement allows one to use the chopper data to set the angular position of the flywheel to $\theta=0$ when the gas volume is a maximum.

\begin{figure}[t]
\centering{
\includegraphics[keepaspectratio, width=0.6\columnwidth]{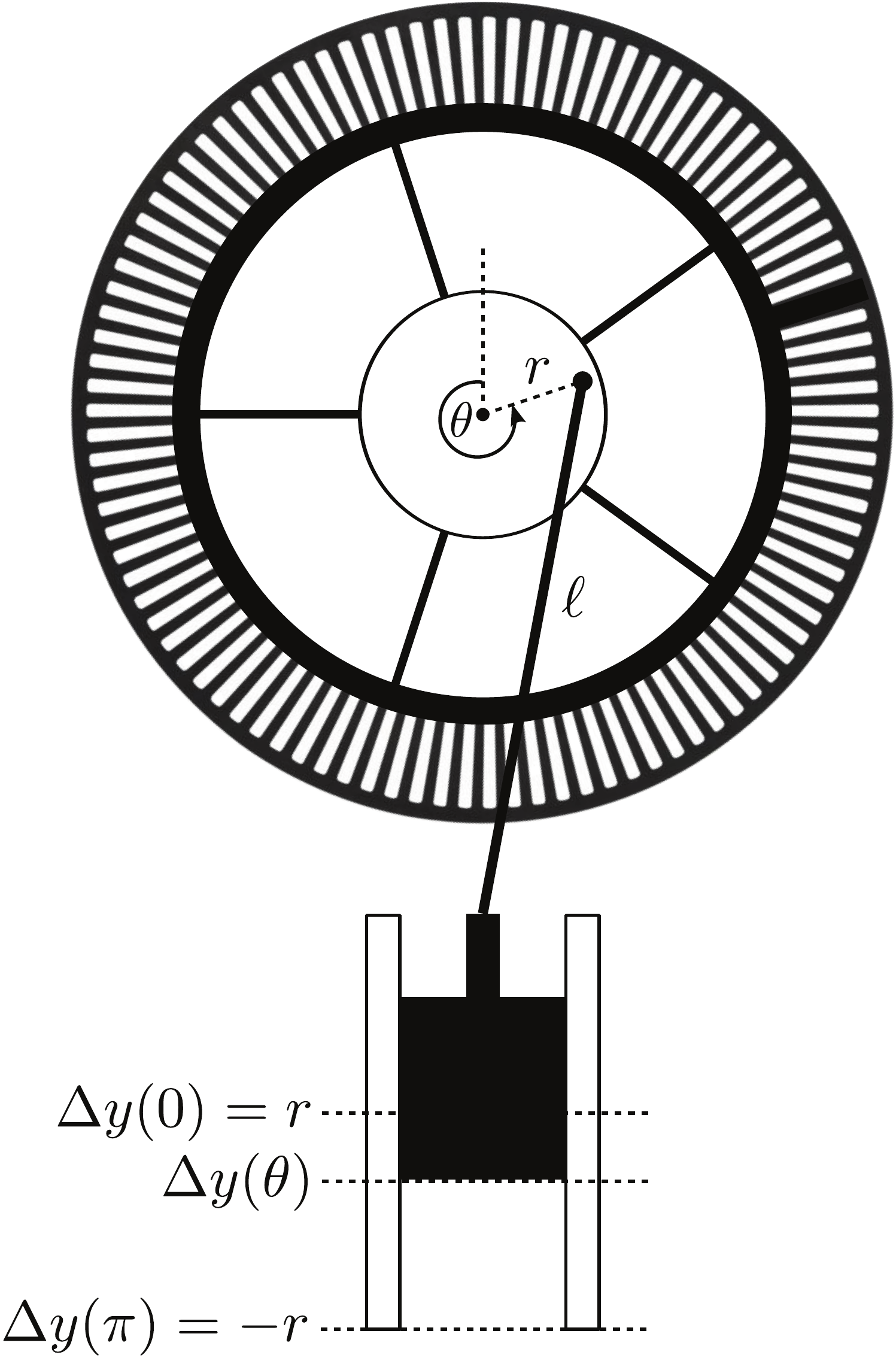}}
\caption{\label{fig:height}The geometry used to convert flywheel position in radians to piston height $\Delta y$.}
\end{figure}

Figure~\ref{fig:height} shows the geometry that relates the flywheel rotation to the piston height.  A careful analysis of this geometry reveals that the change in piston height is given by:
\begin{align}
\Delta y&=r\left[\cos\theta-\frac{\ell}{r}\left(\sqrt{1-\left(\frac{r}{\ell}\sin\theta\right)^2}-1\right)\right]\\
&\approx r\left[\cos\theta+\frac{r}{2\ell}\sin^2\theta\right]
\end{align}  
where the approximate expression is valid when $r\ll\ell$.  The change in volume is then given by $\Delta y$ times the cross-sectional area of the piston.

\begin{figure*}
\centering{
(a)\includegraphics[keepaspectratio, width=0.94\columnwidth]{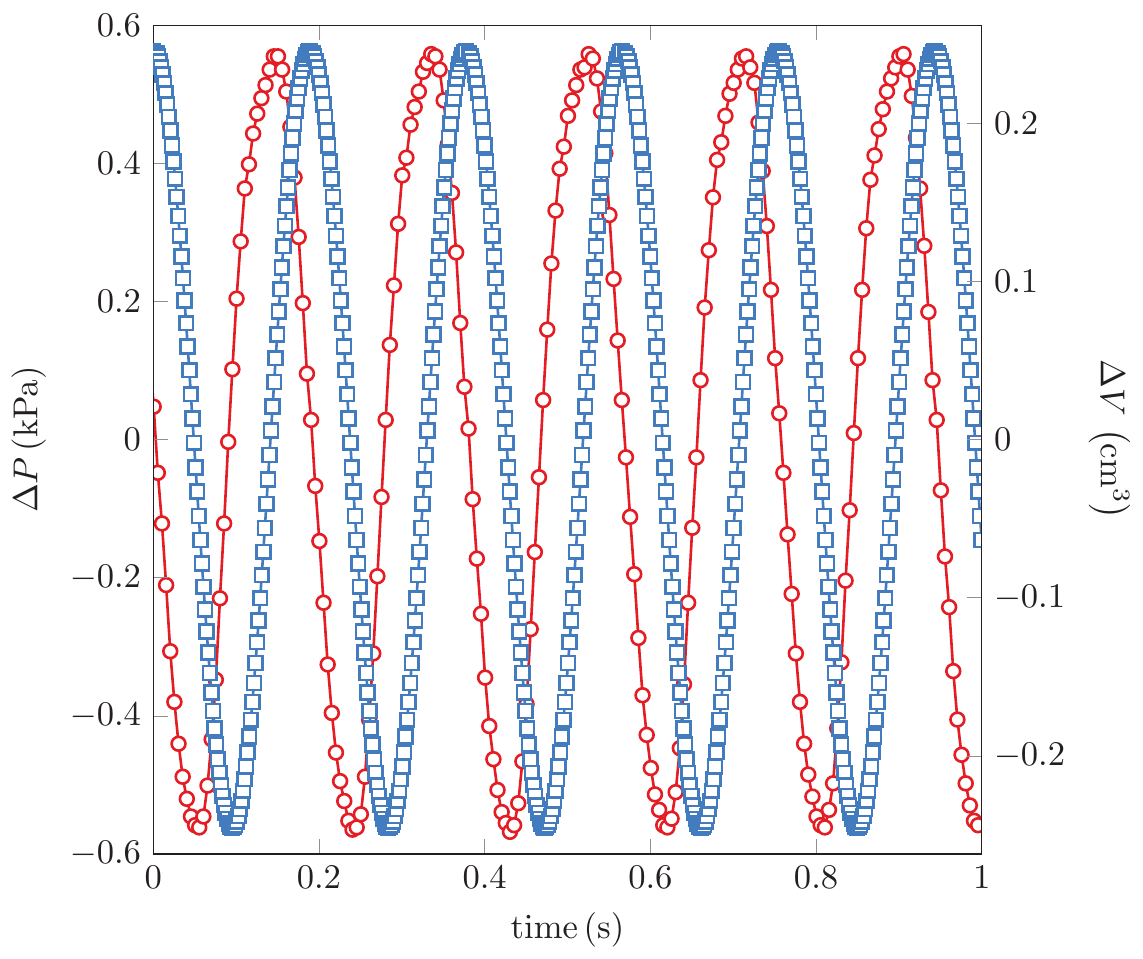}\qquad (b)\includegraphics[keepaspectratio, width=0.94\columnwidth]{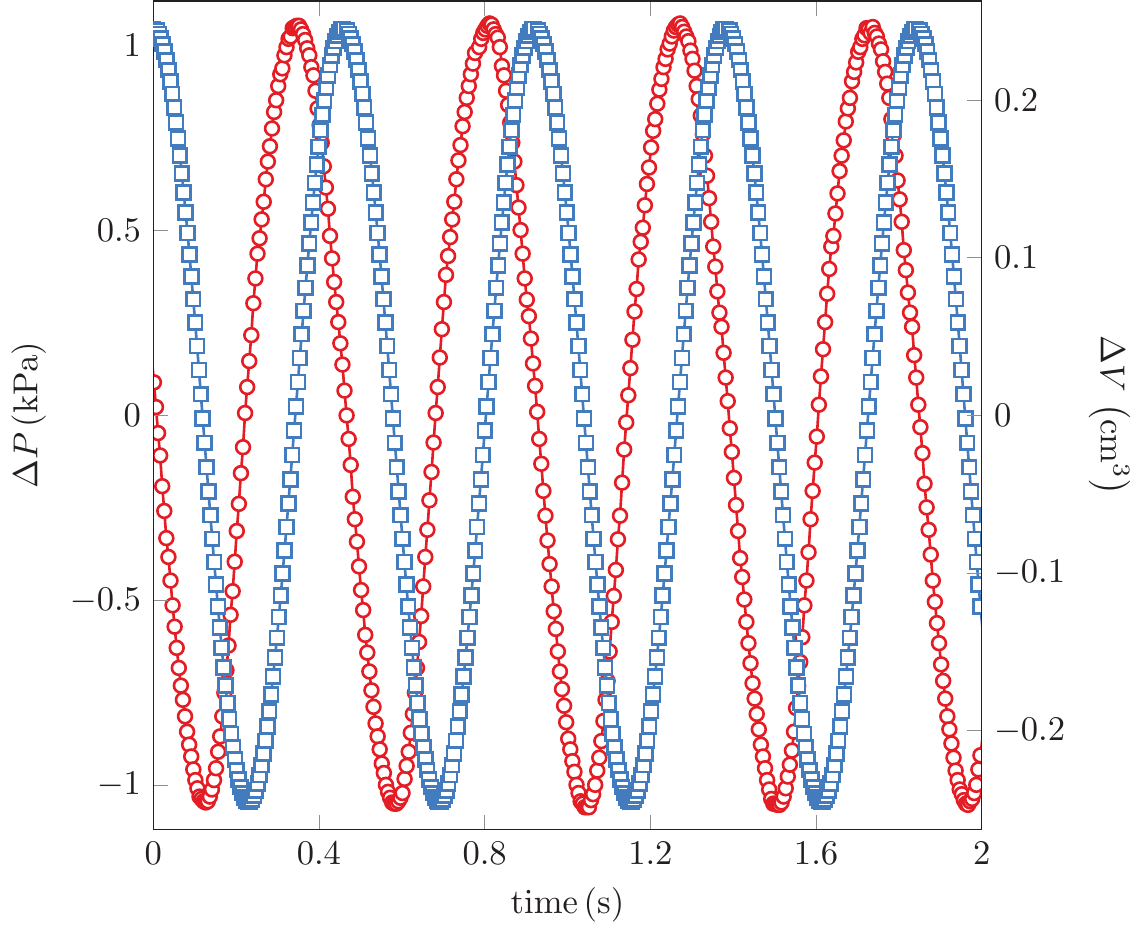}}
\caption{\label{fig:PVvsTime}(a) Measured change in pressure (red circles) and volume (blue squares) versus time with a free-running flywheel.  (b) Measured change in pressure (red circles) and volume (blue squares) versus time with light friction applied to the flywheel.}
\end{figure*}

The angular position of the flywheel was monitored using a Vernier photogate and Logger Pro.  The software was set to record the times at which the photogate went from a blocked to an unblocked state.  Under these conditions, after passing through the $\theta=0$ point, the angle at the $n^\mathrm{th}$ recorded time is given by:
\begin{equation}
\theta_n=\frac{2\pi}{N_0}\left(n-\frac{1}{4}\right)
\end{equation} 
where $N_0=100$ is the number of chopper intervals (before blocking one of the transparent sections), and $n$ is an integer that runs from 1 to $N_0-1$.

\subsection{Results}\label{sec:PVresults}
Figure~\ref{fig:PVvsTime}(a) shows the measured pressure and volume change of the Kontax Stirling engine as a function of time.  For these data, the Stirling engine was placed on top of a cup of hot water.  The water was boiled in a kettle before it was poured into the cup.  The data acquisition was started after the Stirling engine reached a speed that was approximately constant (the speed of the engine slowly decreases as the hot water cools).  The data are high resolution and the acquisition rate is high compared to speed of the engine ($\approx\SI{5.3}{cycles/\second}$).  The data also convincingly show that the pressure and volume changes are \SI{90}{\degree} out of phase and that the pressure leads the volume.

\begin{figure}[t]
\centering{\includegraphics[width=0.94\columnwidth]{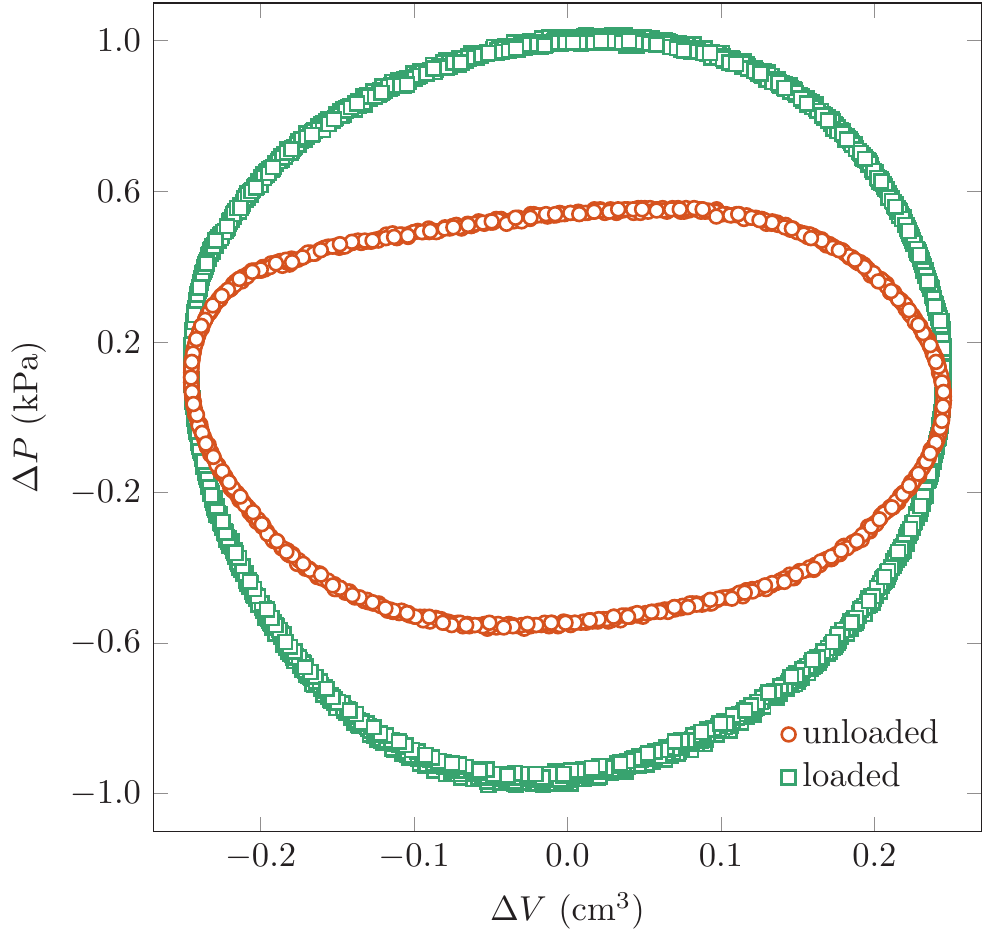}}
\caption{\label{fig:PV}Measured $P$-$V$ diagram of the Kontax Stirling engine with (green squares) and without (orange circles) friction applied to the flywheel.  The added friction causes the engine to do more work per cycle while simultaneously reducing the power output.}
\end{figure}

In Fig.~\ref{fig:PV}, we show a plot of Stirling engine's $P$-$V$ diagram.  The $P$-$V$ diagram that corresponds to the data in Fig.~\ref{fig:PVvsTime}(a) is shown using orange circles.  
Because Logger Pro records the pressure and photogate readings at different times, the mathematics software Maple was used to interpolate the volume changes at the times of the pressure readings.  The plot in Fig.~\ref{fig:PV} shows a set of 25 consecutive cycles that are indistinguishable from one another.

Clearly, the measured $P$-$V$ diagram differs from that of the ideal Stirling cycle.  For truly isothermal processes, one would expect \mbox{$\Delta P\propto \Delta V^{-1}$}.  Also, as expected from the design of the connecting rods, there are no prefectly isochoric processes in the measured cycle.  Nevertheless, the high-resolution data do allow for an accurate measurement of the area enclosed by the loop which gives the work output by the engine per cycle.  Maple was also used perform this calculation.  The work done from measurement $i$ to measurement \mbox{$i+1$} is given by:
\begin{equation}
W_i=\left(\frac{\Delta P_{i+1}+\Delta P_i}{2}\right)\left(\Delta V_{i+1}-\Delta V_i\right)
\end{equation}
Summing all of the $W_i$ contributions from one complete cycle gives the net work.  The result of this calculation was a net work of \SI{0.43}{\milli\joule}.  Multiplying the work by the previously-determined cycle rate of \SI{5.3}{\hertz} gives an output power of \SI{2.3}{\milli\watt}, or $3.0$~micro-horsepower.  A typical car might have a \SI{120}{horsepower} engine -- a factor of 40 million greater than our demonstration Stirling engine!

Finally, the measurements of $\Delta P$ and $\Delta V$ were repeated after light friction was applied to the flywheel of the Stirling engine.  A small piece of sponge was attached to a retort stand and then brought into light contact with the flywheel while the engine was running.  The sponge position was adjusted until the equilibrium period of the engine's cycle noticeably increased.  The pressure and volume changes are plotted as a function of time in Fig.~\ref{fig:PVvsTime}(b).  These data show the same \SI{90}{\degree} phase shift between the pressure and volume.  Notice, however, that the magnitude of the pressure change has approximately doubled while the period of the cycle has increased from \SI{0.19}{\second} for the free-running engine to \SI{0.43}{\second} after applying friction.

The $P$-$V$ diagram that corresponds to the data in Fig.~\ref{fig:PVvsTime}(b) is shown using green squares in Fig.~\ref{fig:PV}.  Again, 25 consecutive and nearly-identical cycles are shown.   The increase in enclosed area is an indication that the engine is doing more work per cycle when friction is applied to the flywheel.  The calculated area is \SI{0.75}{\milli\joule}.  Dividing the work per cycle by the period gives a power output of \SI{1.7}{\milli\watt} or $2.3$~micro-horsepower.  Although loading the flywheel causes the engine to do more work per cycle, the longer period results in power output that, in this case, was reduced by about \SI{20}{\percent}.

\section{Refrigeration}\label{sec:fridge}
Today, the most common commercial application of the Stirling cycle is in closed-cylce cryocoolers.\cite{Koler:1965}  In a cryocooler, mechanical work is done on the working gas to generate a temperature difference. Several companies now make single-stage Striling cryocoolers capable of reaching \SI{40}{\kelvin} and two-stage systems capable of reaching lower temperatures.  For a review of modern cryocooler technology, see Ref.~\onlinecite{Radebaugh:2009}.

\begin{figure}[t]
\centering{
\includegraphics[keepaspectratio, width=\columnwidth]{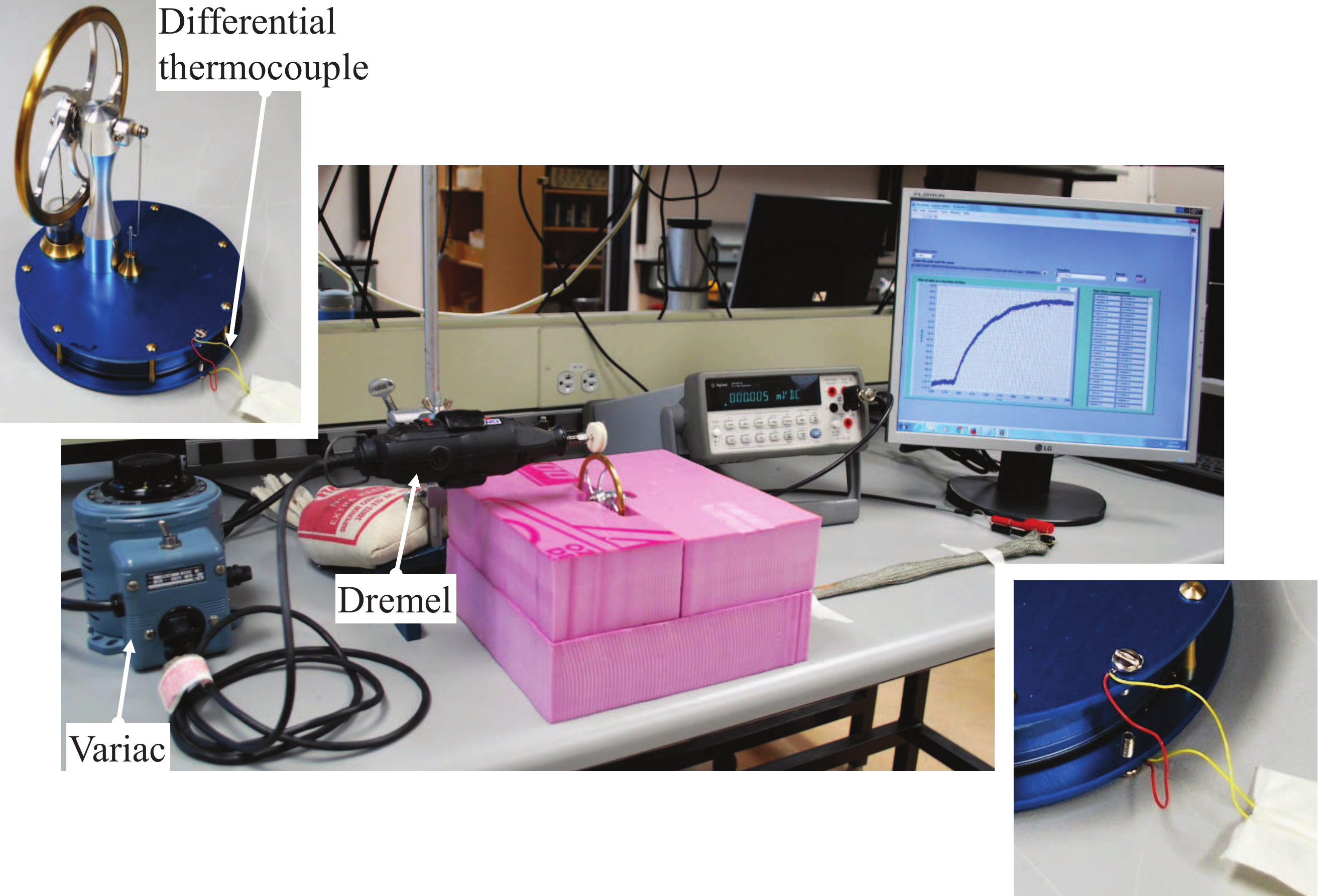}}
\caption{\label{fig:CryocoolerApparatus}Photograph of the setup used to operate the Kontax Stirling engine as a refrigerator.  A Dremel tool was used to rotate the flywheel and a differential thermocouple measured the temperature difference between the top and bottom plates of the engine.}
\end{figure}
To run the Kontax Stirling engine as a cryocooler, it was first necessary to thermally isolate the top and bottom plates of the engine from the surrounding environment.  A \SI[number-unit-product=\text{-}]{2}{inch} sheet of high-density polystyrene foam was cut using a hot \SI[number-unit-product=\text{-}]{32}{AWG} nichrome wire.  The wire was under tension and we found that a current of \SI{1}{\ampere} was sufficient to make make clean cuts.  As shown in Fig.~\ref{fig:CryocoolerApparatus}, polystyrene blocks were cut such that the Stirling engine was completely enclosed except for the upper half of the flywheel. 

The lower-right corner of Fig.~\ref{fig:CryocoolerApparatus} shows the differential thermocouple that was used to monitor the temperature difference between the top and bottom plates of the Stirling engine.  A single \#4-40 tapped hole was drilled into each of the top and bottom plates and screws were used to clamp down on the junctions of a type-E thermocouple.  Fine thermocouple wires were used so as to limit heat conduction along the wires as much as possible.  The thermocouple voltage was measured using a Keysight 34401A multimeter.  The data were logged as a function of time using a simple LabVIEW program.

\begin{figure}[t]
\centering{\includegraphics[width=0.94\columnwidth]{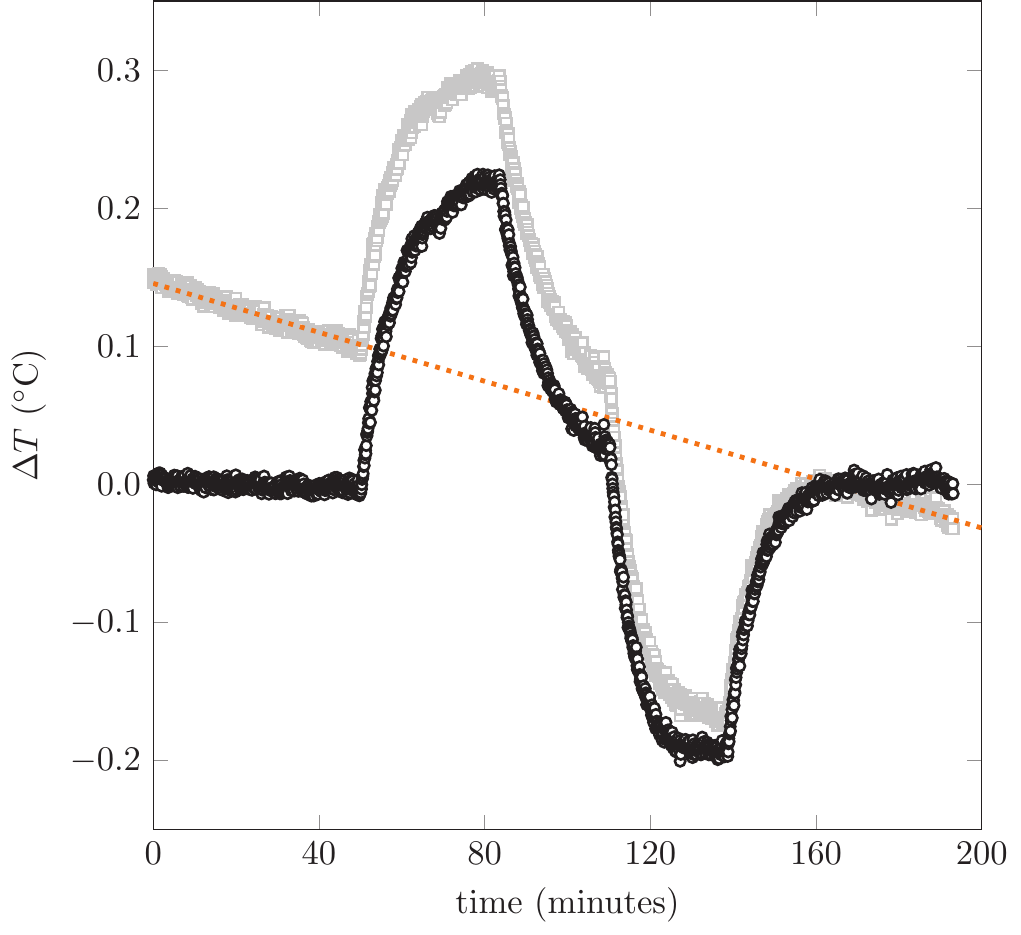}}
\caption{\label{fig:dT}Measured temperature difference across the top and bottom plates of the Stirling engine versus time.  The raw $\Delta T$ data are shown by grey data points.  The black data points show $\Delta T$ after removing the background drift of the room temperature, shown by the dotted line.  A positive $\Delta T$ indicates that the top plate is at a higher temperature than the bottom plate.}
\end{figure}

The flywheel of the Stirling engine was rotated using a Dremel tool and a polishing felt attachment.  Even when on its lowest setting, the Dremel spins the flywheel too fast to produce reliable results.  We found that when the flywheel is rotated too fast, the dominate effect is heating of the top plate by friction.  The best results were obtained when the flywheel was rotated at a rate between \SI{1}{} and \SI{2}{\hertz}.  To reduce the speed of the Dremel, it was plugged into a variac which was set to approximately \SI{25}{Vrms}.

Figure~\ref{fig:dT} shows the difference in temperature between the top and bottom plates of the Stirling engine as a function to time.  The raw data are shown by the grey data points.  Initially, the Dremel was off and the flywheel was stationary.  The Dremel was started at approximately \SI{50}{\minute} and allowed to run continuously for \SI{30}{min}.  During that time, the temperature difference increased and was beginning to plateau by the end of the \SI{30}{\minute} run time.  An increasing $\Delta T$ corresponds to the top plate warming relative to the bottom plate.  After the Dremel was stopped, the temperature difference steadily decreased.  The Dremel was run for another \SI{30}{\minute} interval starting at approximately \SI{110}{\minute}.  However, for this run the rotation direction of the flywheel was reversed.  As expected, this caused the top plate to cool relative to the bottom plate.  The Dremel was stopped at \SI{140}{\minute}.

Figure~\ref{fig:dT} clearly shows a nonzero $\Delta T$ that evolves linearly with time even when the Dremel has been stopped for an extended period of time.  This is due to a slow drift of the room temperature.  The polystyrene used to isolate the Stirling engine from its surroundings has a hole in the top to accommodate the flywheel.  Therefore, the top plate of the engine is more sensitive to changes in room temperature than the bottom plate which results in the observed drift in $\Delta T$.  The grey data from \SI{0}{} to \SI{48}{\minute} and from \SI{160}{} to \SI{190}{\minute} were fit to a straight line.  The best fit line has a slope of \mbox{\SI{-0.89e-3}{\celsius/\minute}} and is shown in Fig.~\ref{fig:dT} as the dotted orange line.  The black data in the figure show $\Delta T$ after removing the drift caused by the changing room temperature.  The result is an approximately symmetric $\Delta T$ that reaches \SI{\pm 0.2}{\celsius} after running the Dremel for about \SI{30}{\minute}.

\section{Summary}\label{sec:summary}
Stirling engines are commonly used as a demonstration of a working heat engine.  These demonstrations can be significantly enhanced by adding instrumentation to measure the pressure and volume changes of a running engine.  We have described a simple and inexpensive way to instrument a high-quality Stirling engine designed for demonstrations.  Our pressure versus time measurements are high resolution and reveal the \SI{90}{\degree} phase shift between the pressure and volume changes.  The $P$-$V$ diagram constructed from these measurements allows one to determine the net work by the engine per cycle and, therefore, the power output.  By adding friction to the flywheel, it was shown that the work per cycle increased while the power output decreased.  Finally, we used a rotary tool to drive the flywheel of the Stirling engine while monitoring the temperature difference across its body.  A small but measurable temperature difference was observed.  Reversing the flywheel rotation direction reversed the sign of the observed temperature difference.

\appendix

\section{Parts and Suppliers}\label{app:parts}

This appendix provides a list of the equipment required to reproduce the measurements described in this paper.  Where appropriate, vendors and cost estimates are also provided.  

{\it Stirling Engine} --  The Stirling engine used was manufactured by Kontax Stirling Engines \mbox{(\url{https://www.stirlingengine.co.uk/})}.  We purchased the KS90 Blue model which costs \$118. 

\subsection{P-V diagram}

{\it Data Acquisition} -- Our data was acquired using Vernier's differential voltage probe (\$54), photogate (\$62), LabQuest interface (\$203), and Logger Pro software (\$339) \mbox{(\url{https://www.vernier.com/})}.  These items are standard in many university and college physics departments.  We also note that the data acquisition could also be implemented using alternative systems, like those offered by PASCO \mbox{(\url{https://www.pasco.com/})} and National Instruments \mbox{(\url{http://www.ni.com/})}.   

{\it Pressure Sensor} -- The MPXV7002 pressure sensor made by NXP Semiconductors was purchased from Digi-Key for \$19 \mbox{(\url{https://www.digikey.com/})}.  The pressure sensor requires \SI{5}{\volt} DC to operate.  We used the E3630A power supply from Keysight, but any DC power supply can be used. 

{\it Miscellaneous} -- Also required are a kettle to boil water, a coffee mug, a syringe, epoxy (J-B Weld is a good and inexpensive option), and rubber tubing.  Wires also need to be soldered to the pressure sensor.

\begin{figure*}
\centering{\includegraphics{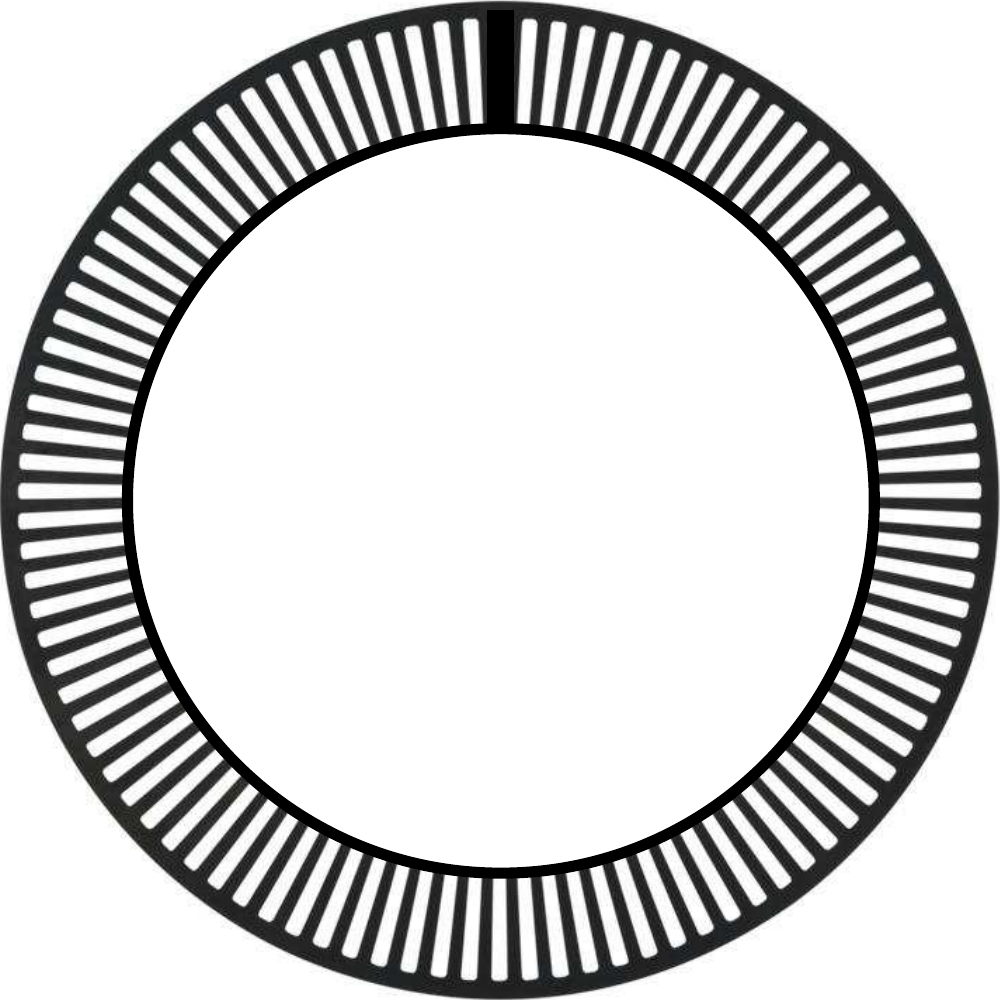}}
\caption{\label{fig:chopper}The chopper design used to monitor the flywheel rotation.  The outside diameter of the chopper is \SI{4}{inches}.}
\end{figure*}

\subsection{Refrigeration}

{\it Rotary Tool} -- We used a Dremel tool that can be purchased from most hardware stores for approximately \$50.  We had to reduce the supply voltage to the Dremel using a variable transformer (variac).  Many physics departments will already have such an instrument.  They can, however, be purchased from many suppliers.  For example, they are available from Digi-Key for \$370 \mbox{(\url{https://www.digikey.com/})}.

{\it Digital Multimeter} -- The thermocouple voltage is very small and the Vernier voltage probe does not have the required sensitivity.  We measured the thermocouple voltage using Keysight's 34401A multimeter which is no longer in production.  Their replacement is the 34461A multimeter which costs \$1121 \mbox{(\url{https://www.keysight.com/})}.  The multimeter we used was interfaced to a computer running LabVIEW via a GPIB-to-USB converter made by National Instruments.  One advantage of the new 34461A multimeter is that it can connect to a computer directly via USB.

{\it Miscellaneous} -- This measurement also requires thermocouple wires which can be purchased from Omega \mbox{(\url{https://www.omega.com/})} and polystyrene insulation which can be found at most hardware stores.

\section{Chopper Design}\label{app:chopper}

The chopper that was used to track the rotation of the Stirling engine's flywheel is shown in Fig.~\ref{fig:chopper}.  As described earlier, the thick dark band at the top was used to indicate that the piston was at its peak height.  In our work, this image was printed onto an acetate sheet (overhead transparency) and then cut out along the inside and outside circumferences.  The chopper was then taped to the flywheel of the Stirling engine using a few small pieces of cellophane tape.  If printed without any scaling, the image in Fig.~\ref{fig:chopper} will fit onto the flywheel of the KS90 models of the Stirling engines manufactured by Kontax.

\end{document}